\documentclass[pre,preprintnumbers,amsmath,amssymb]{revtex4}
\usepackage{graphicx}% Include figure files
\usepackage{bm}% bold math

\begin{document}
\title{Maximum of $N$ Independent Brownian Walkers till the First Exit From
the Half Space}

\author{P.~L. Krapivsky$^1$, Satya N. Majumdar$^2$, Alberto Rosso$^2$}
\affiliation{
$^1$ Department of Physics, Boston University, Boston, MA 02215, USA \\
$^2$ Univ. Paris Sud, CNRS, LPTMS, 
UMR 8626, Orsay F-91405, France}

\begin{abstract}
We consider the one-dimensional target search process that involves an 
immobile target located at the origin and $N$ searchers performing independent Brownian motions starting at the initial positions $\vec x = (x_1,x_2,\ldots, x_N)$ all on the positive half space. The process stops  when the target is first found by one of the searchers. We compute the probability distribution of the maximum
distance $m$ visited by the searchers till the stopping time
and show that it has a power law tail: $P_N(m|\vec x)\simeq B_N (x_1x_2\ldots 
x_N)/m^{N+1}$ for large $m$. Thus all moments of $m$ up to
the order $(N-1)$ are finite, while the higher moments diverge. The prefactor 
$B_N$ increases with $N$ faster
than exponentially. Our solution gives the exit probability of a set of $N$ particles from a box $[0,L]$ through the left boundary. Incidentally, it also provides
an exact solution of the Laplace's equation in an $N$-dimensional
hypercube with some prescribed boundary conditions. The analytical
results are in excellent agreement with Monte Carlo simulations.
  
\end{abstract}

\maketitle

\date{\today}

\section{introduction}

The probability distribution of the maximum of a single one-dimensional Brownian 
motion (and its variants such as a Brownian bridge or an excursion) over a fixed
interval of time $[0,t]$, has a long history in the probability 
literature~\cite{Chung,Kennedy,Durret,Darling,PY,Finch,MRKY}.
The statistics of the maximum has diverse applications. One example is the
Kolmogorov-Smirnov test in statistics that is used to compare, in
a nonparametric way, two different probability distributions~\cite{KS1,KS2}.
Similarly, the distribution of the global maximum of a discrete-time random 
flights (including L\'evy flights) has also been studied
in the probability literature~\cite{Pollaczek,Spitzer}, with more
recent applications in computer science~\cite{Flajolet}, 
physics~\cite{rwmax} and chemistry~\cite{Ziff}.

In statistical physics, there has been a recent revival of 
interest in related problems in the context of the distribution 
of the maximal height, measured with respect to a reference point, of $(1+1)$-dimensional fluctuating interfaces~\cite{Raychow,MC,interface}.
In the stationary state of a finite sample of size $L$, such fluctuating 
interfaces are often described by a Brownian bridge in space over an interval 
$[0,L]$, albeit with certain global constraints~\cite{MC}. 
The statistics of maximum has also been computed for
continuous-time
subdiffusive processes~\cite{SL,Barkai} and 
has been used to analyse single particle trajectories~\cite{Tejedor}.

The distribution of the maximum for a 
single Brownian motion 
(or its variants such as bridge, excursion etc.) 
has been extended to many Brownian motions, 
including certain strongly interacting random walkers, 
e.g. non-intersecting, so-called vicious random 
walkers~\cite{SMCR,Katori,Feierl,NM}. 
(The latter problem has an intriguing connection to the 
Gaussian ensembles of the random matrix 
theory~\cite{J,SMCR,NM}.) For independent walkers, the results 
on the distribution of the maximum have recently been used
to compute the mean perimeter and the mean area of the
convex hull of $N$ independent planar Brownian motions~\cite{RMC,chull}.   
 
\begin{figure}
\includegraphics[height=8.0cm,width=8.0cm,angle=0]{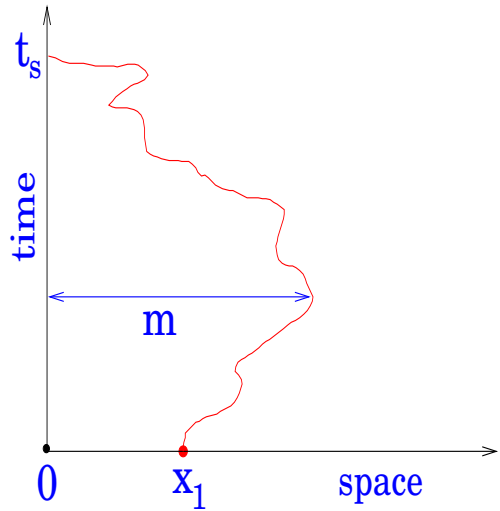}
\caption{\label{fig:n1max} The trajectory (red line) of a single Brownian 
walker
starting initially at $x_1$ till the stopping time $t_s$ at which it 
hits the origin for the first time. The maximum distance travelled
by the particle till $t_s$ is denoted by $m$.}
\end{figure}

These results on Brownian motion and its variants represent
rare exact analytical results for the extreme value statistics
of {\em correlated} random variables, a subject of increasing
current interest~\cite{MK}.
However, all these results about the distribution of the maximum, for a single 
or multiple walkers, have been derived
in the case when one considers the walkers over a fixed interval of time 
$[0,t]$. An interesting variation of this problem, with several applications,
arises when the interval $[0,t]$ is not fixed, but itself varies
from realization to realization, i.e., one observes
the walker (or walkers) over a time interval $[0,t_s]$ where the
stopping time $t_s$ of the process itself is a random variable. For example,
$t_s$ may represent the first-passage time (through the origin)
of a walker. 

To be more precise, consider first a  single Brownian walker that starts at time $t=0$ at position $x_1>0$. The position $x_1(t)$ of the walker evolves via the continuous-time stochastic equation, $dx_1/dt=\eta_1(t)$ where $\eta_1(t)$ is
a Gaussian white noise with mean $\langle \eta_1(t)\rangle =0$
and a correlator $\langle \eta_1(t)\eta_1(t')\rangle= \delta(t-t')$.
The process stops at the stopping time $t_s$ when the walker hits the origin 
for the first time (see Fig.~\ref{fig:n1max}). 

Let $m$ be the maximum displacement of the particle till the stopping time $t_s$. 
The statistics of the random variable $m$ is interesting and
it represents an example of the so called first-passage
Brownian functional~\cite{review}. The problem is a toy model of `random search', where the origin represents a fixed `target' and the Brownian walker
represents a random searcher. The search is called off when the searcher finds 
its target and $m$ represents the maximum distance travelled by the searcher
before it finds its target. For concreteness, we shall mostly used terminology related to random search, although there are several applications
of this problem. For example, in the context of trapping~\cite{TA,BZK,BO} or 
predator-prey~\cite{RK}  models the origin may represent an immobile target (prey) and the Brownian walker may represent a diffusing chemical trap (predator).
The stopping time $t_s$ is then the reaction time or the survival
time of the prey and $m$ denotes the maximum distance the predator
travels before finding its prey. In the context of the directed 
Abelian sandpile model in $(1+1)$-dimensions~\cite{DR}, $m$ represents the 
maximum lateral size of an avalanche~\cite{KM1}. The random variable $m$
also plays an important role in characterizing 
the so called staircase polygons~\cite{Kearney}. In the context of queueing 
theory, where the position of the walker represents the length of a queue,
$m$ represents the maximum length of a queue during the so called busy 
period~\cite{Kearney,KM1}.

The probability density function (pdf) $P_1(m\big|x_1)$ of $m$ (for fixed 
$x_1$) can be easily computed~\cite{KM1} and it turns out to be a pure 
power law:
\begin{equation}
P_1(m\big|x_1)=\frac{x_1}{m^2};\quad\, m\ge x_1.
\label{n1pdf0}
\end{equation}
While this pdf is evidently normalized to unity, the average $\langle m\rangle$
and higher integer moments are infinite! The cumulative distribution of the maximum is given by
\begin{equation}
Q_1(L\big|x_1)={\rm Prob}\left[m\le L\big|x_1\right]= \int_{x_1}^L 
P_1(m\big|x_1)\, 
dm=1-\frac{x_1}{L}\,.
\label{n1cum}
\end{equation}
This distribution has a very simple interpretation: it just represents the exit
probability of a Brownian particle~\cite{Feller}, starting at $0\le x_1 \le L$, 
from a box $[0,L]$ through its left boundary at $0$.

\begin{figure}
\includegraphics[height=8.0cm,width=8.0cm,angle=0]{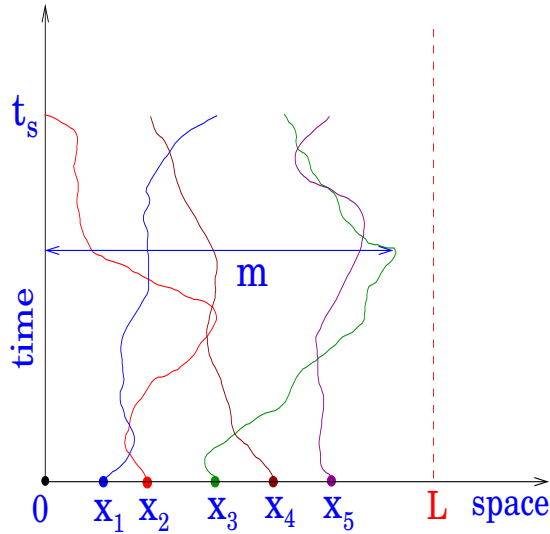}
\caption{\label{fig:mult} 
The trajectories of $N=5$ independent Brownian 
walkers starting at initial positions $x_1$, $x_2$, $x_3$, $x_4$ and $x_5$
till the stopping time $t_s$ when one of the walkers
(the second one (red) in this figure) hits the origin.
The maximum displacement along the $x$ direction till $t_s$ (undergone, e.g.,
by the third particle (green)) is denoted by $m$. The cumulative probability
$Q_N(L\big|\vec x)={\rm Prob}\left[m\le L|\vec x\right]$ also represents
the exit probability of the first particle from a box $[0,L]$ through
its left boundary.}
\end{figure}

In this paper, we study a generalization of this search
problem where there is still one fixed target at the origin, but
there are $N$ searchers who perform independent Brownian
motions on the $x>0$ axis, starting at the initial positions ${\vec x}\equiv 
(x_1,x_2,\ldots, x_N)$. 
The position $x_i(t)$ of the $i$-th walker evolves with time $t$ via the  
Brownian evolution
\begin{equation}
\frac{dx_i}{dt}= \eta_i(t)
\label{brown1}
\end{equation}
where $\eta_i(t)$ is a Gaussian white noise satisfying $\langle \eta_i(t)\rangle =0$ and $\langle \eta_i(t)\eta_j(t')\rangle= \delta(t-t')\delta_{i,j}$. 
Since the walkers are independent they can cross each other.
The process stops at a stopping time $t_s$ when the origin is hit for the 
first time by any one of the walkers (e.g., the second walker (red) in 
Fig. \ref{fig:mult}). Note that $t_s$ varies from 
one history of the process to another. 

In the context of chemical kinetics~\cite{BZK} where the problem is generally referred to as the `target annihilation' problem, various generalizations 
of this problem have been investigated including e.g. the situation where the target itself diffuses~\cite{BL,BB,Oshanin,BMB,Yuste}. In the following, we shall limit
ourselves to the case of immobile target and focus on the statistics of 
the maximum distance $m$ (from the target) travelled by
any of the walkers till the stopping time $t_s$ when the target is found. 
Thus $m$ denotes the distance
of the farthest point on the $x$ axis visited by any one of the walkers
till $t_s$. Clearly $m$ is a random variable fluctuating from one realization
of the process to another. Our object of interest is the probability
density $P_N(m\big|{\vec x})$ of this maximum distance $m$, given the number 
$N$ of walkers and their initial positions $\vec x$. 
Thus $m$ provides an estimate (worst-case) of the distance
that needs to be covered by a team of $N$ walkers to find a fixed target.

As in the single searcher case, 
let $Q_N(L\big|{\vec x})={\rm Prob}[m\le L\big|{\vec 
x}]=\int_{0}^{L}P_N(m\big|\vec x)\,dm$ be the cumulative 
probability that the
maximum $m$ till $t_s$ is less than or equal to $L$. 
This cumulative distribution of the maximum can be interpreted
as the solution of a different problem as in the $N=1$ case.
Consider, for instance,
a slightly different problem where again we have a set of 
$N$ independent walkers, but now inside a box $[0,L]$,
starting at the initial positions $\vec x$. Let us define the
exit probability as the probability that the first particle that exits the
box $[0,L]$ does so through $0$ (and not through the upper boundary at $L$), 
see Fig. \ref{fig:mult}. As in the $N=1$ case, this exit probability
is precisely the cumulative distribution $Q_N(L\big|\vec x)$ of the maximum 
$m$ till the stopping time in the semi-infinite system, as it counts 
all those events where one of the trajectories hits the lower boundary $0$
before hitting the upper boundary at $L$ while all the others
stay inside the box $[0,L]$ till this event 
of first-hitting the origin.  

We will see that for this seemingly simple one-dimensional model of 
independent walkers, the statistics of $m$ has a rich
and nontrivial dependence on the number $N$ of walkers.
This is partly due to the fact that the same stopping time $t_s$
for all the walkers effectively introduces a correlation between the
trajectories of the walkers,
even though each executes an independent Brownian motion.
While for $N=1$ the solution is simple, it becomes rather
nonrivial even for $N=2$!

Let us first summarize our main results. We compute the pdf $P_N(m\big|\vec x)$ 
exactly for all $N\ge 1$ by a path counting (or path integral) method.
We show that, for arbitrary $N\ge 1$, the pdf of the maximum has an asymptotic 
power-law tail
\begin{equation}
P_N(m\big|{\vec x})\simeq B_N\, \frac{x_1\,x_2\,\ldots x_N}{m^{N+1}}\quad 
{\rm 
as}\,\, m\to \infty
\label{max1}
\end{equation}
where the prefactor $B_N$ has a nontrivial $N$ dependence which
we compute explicitly. For $N=1$, we have $B_1=1$ and the asymptotic result 
in \eqref{max1} is actually valid exactly for all $m\geq x_1$. For $N=2$, we
will see that
\begin{equation}
B_2= 
\frac{1}{4\pi^2}\left[\Gamma\left(\frac{1}{4}\right)\right]^{4}=4.37688\ldots
\label{amp2}
\end{equation}
In particular, for large $N$, the prefactor $B_N$ grows faster than
exponentially
\begin{equation}
B_N \simeq N\left[\frac{4}{\pi}\ln(N)\right]^{N/2}.
\label{ampN}
\end{equation}
Our asymptotic result \eqref{max1}
indicates that for $N$ walkers, integer moments of $m$ up to order 
$(N-1)$ are finite, while higher integer moments are infinite. Evidently, as $N$ 
increases, the
distribution becomes narrower and narrower as expected but it does
so in a nontrivial fashion.

Recently, the cumulative distribution of the maximum $m$ till the first-passage 
time $t_s$, or equivalently the exit probability $Q(L|x)$ from the box $[0,L]$
through the origin, 
was studied~\cite{MRZ1} for a generic self-affine stochastic process $x(t)$ 
starting at the initial position $x$. The process $x(t)$ typically grows with 
time as $x(t)\sim t^H$ where
$H$ is the Hurst exponent. This power law growth of distance with time
makes the process self-affine. An example is
the ordinary Brownian motion where $H=1/2$.
For such a generic self-affine process, it was
argued~\cite{MRZ1} that the cumulative distribution 
of the maximum $Q(L|x)={\rm Prob}(m\le L|x)\sim 1- A\, (x/L)^{\phi}$ 
in the limit $x/L \to 0$ where $A$ is a constant. The exponent $\phi$ was 
found to be related to the persistence exponent $\theta$ via the scaling
relation $\phi=\theta/H$ ~\cite{MRZ1}. The persistence exponent $\theta$
characterizes the late time power law decay of the survival
probability, i.e., the probability that the process stays
on the positive half-axis up to time $t$~\cite{pers_review}.
Thus the pdf of the maximum decays for large
$m$ as, $P(m|x)\sim m^{-\phi-1}$ with $\phi=\theta/H$.
The exact result \eqref{max1} shows that if we think of the assembly
of $N$ independent Brownian motions as a single self-affine
stochastic process in the $N$-dimensional space,
then $\phi=N$. We will see later that the persistence exponent
for this collective process is $\theta=N/2$ and the Hurst exponent $H=1/2$.
Thus our exact result for this model supports the general scaling relation
$\phi=\theta/H$ found in \cite{MRZ1}.  

The paper is organized as follows. In Section 2, we provide a simple
heuristic argument in favour of our main result \eqref{max1}.
This argument is not sufficient to compute the prefactor $B_N$
exactly for all $N$. However, we show that this heuristic argument
becomes asymptotically exact for large $N$ and one can extract
the limiting behavior of $B_N$ for large $N$ using
an extreme value argument. In Section 3, we set up the general
method for computing the cumulative distribution $Q_N(L\big|\vec x)$ of the
maximum $m$. This requires solving Laplace's equation in an $N$-dimensional 
space with appropriate boundary conditions. We present explicit solutions
for the cases $N=1$ and $N=2$. In Section 4, we present an alternative
path counting method that is more general, physically transparent 
and provides explicit results for all
$N\ge 1$. In Section 5, we present numerical results to verify our analytical   
predictions. Finally in Section 6, we conclude with a summary
and a list of interesting open problems. Some of the 
details of the computations are relegated
to the appendices.

\section{Heuristic Argument}

We begin with a simple heuristic argument in favour
of \eqref{max1}. Consider the semi-infinite geometry with
the independent Brownian motions, evolving via \eqref{brown1}, starting at 
initial positions $\{x_1,x_2,\ldots,x_N\}$. Let $t_s$ be the stopping time
when one of the walkers hits the origin. The probabality distribution
of $t_s$ can be computed exactly as follows. 
As an input to solving the $N$-particle problem, let us first consider
a single Brownian motion starting initially at $x_0>0$
and let $p(x_0,t)$ be the survival probability, i.e., the probability that the
walker does not hit the
origin up to time $t$. This can be easily computed by various standard
methods and is given 
by the well known formula~\cite{redner,review}
\begin{equation}
p(x_0,t)={\rm erf}\left(\frac{x_0}{\sqrt{2t}}\right); \quad {\rm 
erf}(y)=\frac{2}{\sqrt{\pi}}\,\int_0^y e^{-u^2}\,du.
\label{surv1}
\end{equation}

Turning to $N$ walkers we note that since the walkers are independent, the probability that none of them hits $0$ up to time $t$ is simply the product   
$\prod_{i=1}^N p(x_i,t)$. This is precisely the probability
that the stopping time $t_s>t$. Thus, the cumulative distribution
of the stopping time $t_s$, given the initial positions ${\vec x}$, reads~\cite{BZK,BB}
\begin{equation}
{\rm Prob}\left[t_s> t\big|\vec x\right]= \prod_{i=1}^N {\rm 
erf}\left(\frac{x_i}{\sqrt{2t}}\right).
\label{surv2}
\end{equation}
For large $t$, this cumulative distribution has a power law tail
\begin{equation}
{\rm Prob}\left[t_s> t\big|\vec x\right] \simeq 
\left(\frac{2}{\pi}\right)^{N/2}\, 
\frac{x_1\,x_2\,\ldots x_N}{t^{N/2}}\,.
\label{surv3}
\end{equation}
This asymptotic for survival probability tells us that the persistence exponent is 
$\theta=N/2$.

The result \eqref{surv2} and its asymptotic counterpart \eqref{surv3}
are exact. Next comes the heuristic part. We note that for large $t_s$, the
typical maximal displacement $m$ in time $t_s$ must scale as: $m\sim 
\sqrt{t_s}$. Taking this relationship between the two random variables $m$
and $t_s$ seriously, we see that ${\rm Prob}\left[m> L\big|\vec x\right]\sim {\rm 
Prob}\left[t_s>L^2\big|\vec x\right]$ for large $L$. 
Since $Q_N(L\big|\vec x)={\rm Prob}\left[m\le L\big|\vec x\right]$, we conclude that for large $L$
\begin{equation}
1-Q_N(L\big|\vec x)={\rm Prob}[m> L\big|\vec x]\sim {\rm 
Prob}[t_s>L^2\big|\vec 
x]\sim 
\frac{x_1\,x_2\,\ldots, x_N}{L^N}
\label{h1}
\end{equation}
where we used the result in Eq. (\ref{surv3}).
Taking derivatives with respect to $L$ and putting $L=m$ then gives
an approximate behavior of the probability density $P(m\big|\vec x)$
of the maximum for large $m$
\begin{equation}
P_N(m\big|\vec x)\sim \frac{x_1\,x_2\,\ldots, x_N}{m^{N+1}}
\label{max2}
\end{equation}
as announced in \eqref{max1}. This heuristic scaling argument 
thus provides, up to an overall $N$ dependent prefactor $B_N$,
the leading asymptotic power law tail of the distribution of $m$
in \eqref{max1}.

To compute the prefactor $B_N$ exactly for any $N$ one needs
to go beyond this scaling argument. This requires 
a more sophisticated mathematical analysis that is carried out in the
rest of the paper. However, it is possible to
refine this heuristic argument, as shown below, that even
provides the prefactor $B_N$ exactly for large $N$.

In the argument above, the main approximation was to replace
$m\sim \sqrt{t_s}$ for large $m$ and then use the exact asymptotic distribution
of $t_s$ in Eq. (\ref{surv3}) to compute the tail of the distribution of $m$.
This approximation clearly ignores fluctuations of $m$ for a fixed $t_s$.
We now use an extreme value argument to show that this approximation actually 
becomes exact for large $N$.
We consider again a group of Brownian motions starting
at the initial positions $\vec x$ and examine their
trajectories over a {\em fixed} time interval $[0,t_s]$, with
$m$ denoting their global maximum in $[0,t_s]$. 
To compute the cumulative probability $Q_N(L,t_s\big|\vec x)={\rm 
Prob}\left[m\le L\big|\vec x\right]$, we consider the trajectories that stay below
the level $L$ till $t_s$ and also above the level $0$. 
Now, for large $L$, the trajectories that contribute
to $Q_N(L,t_s\big|\vec x)$ typically have large excursions.
So, to a first approximation, one can ignore the lower
boundary at $0$. For the $i$-th walker, starting at $x_i$, the probability
that its maximum stays below $L$ can be computed easily: it is just
the survival probability $p(x_0,t_s)$ in Eq. (\ref{surv1})
with the initial position $x_0=L-x_i$. Thus the joint probability
that all walkers stay below $L$ till $t_s$ (ignoring
the lower boundary at $0$) is just the product
\begin{equation}
Q_N(L,t_s\big|\vec x) \simeq \prod_{i=1}^N {\rm 
erf}\left(\frac{L-x_i}{\sqrt{2t_s}}\right).
\label{heu1}
\end{equation}
For large argument, the error function behaves as $1-{\rm erf}(x)= 
e^{-x^2}/(x\sqrt{\pi})\simeq e^{-x^2}$ to leading order. 
Hence, for large $L$ and large $N$, one can 
write
\begin{equation}
Q_N(L,t_s\big|\vec x)\simeq \exp\left[-N\, e^{-L^2/{2t_s}}\right]\to  
f\left[(L-a_N)/b_N\right]
\label{heu2}
\end{equation}
where the scale factors $a_N= \sqrt{2t_s \ln (N)}$ and 
$b_N=\sqrt{t_s/(2\ln N)}$ and the scaling function $f(x)=\exp[-e^{-x}]$
is the standard Gumbel function. The pdf of $m$ is just the derivative
of the cumulative distribution.
The derivative of the scaling function $f'(x)=\exp\left[-x-e^{-x}\right]$ 
has a peak at $x=0$.
This indicates that the random
variable $m$ has a peak at $m=a_N= \sqrt{2t_s \ln (N)}$ and the width
of $m$ around its peak is $b_N=\sqrt{t_s/(2\ln N)}$ that actually
decreases with increasing $N$. Thus, for large $N$, the random
variable $m$ approaches to the its mean value $a_N$ with
probability $1$, i.e., $m=\sqrt{2t_s \ln (N)}$ with fluctuations
around this value essentially negligible for large $N$.
Using this relation 
in Eq. (\ref{surv3})
provides the following tail for the cumulative distribution of $m$
for large $N$
\begin{equation}
1-Q_N(L\big|\vec x)={\rm Prob}\left[m> L\big|\,\vec x\right]\simeq {\rm 
Prob}\left[t_s> 
\frac{L^2}{2} 
(\ln 
N)^{-2}\big|\, \vec x\right]\simeq A_N \frac{x_1\,x_2\,\ldots, x_N}{L^N}
\label{heu3}
\end{equation}
with the prefactor $A_N\simeq \left[4\ln(N)/\pi\right]^{N/2}$. Taking 
derivative 
with respect to $L$ then gives the tail of the pdf of $m$ in Eq. (\ref{max1})
with the prefactor, for large $N$,
\begin{equation}
B_N= N A_N \simeq N\left[\frac{4}{\pi}\ln(N)\right]^{N/2}.
\label{ampNheu} 
\end{equation} 
We will see later that the same asymptotic result also follows from a
more rigorous approach.

\section{A backward Fokker-Planck method: Laplace's Equation}
\label{Lap-sec}

In this section we show that the cumulative distribution $Q_N(L\big |\,\vec 
x)={\rm Prob}\left[m\le L\big|\,\vec x\right]$ of the maximum $m$ satisfies a 
Laplace's equation in an $N$-dimensional hypercube with appropriate boundary conditions. To see this, it is first useful to consider $Q_N(L\big|\vec x)\equiv Q_N(\vec x\big|L)$ as a function of the coordinates $\vec x$ for a given fixed $L$.
Note that $Q_N(\vec x\big|L)$ is the probability that starting at
$\vec x$, the maximum of the process till $t_s$ stays below the
level $L$. The idea is to derive a differential equation for $Q_N(\vec x\big|L)$ 
using a backward approach where one focuses on the evolution
of the system via \eqref{brown1}
over a small time interval $[0,dt]$ starting from
the initial positions $\vec x$. According to \eqref{brown1}, in this small time 
interval $dt$, the $i$-th particle moves from $x_i$ to ${x_i}'=x_i+\eta_i(0)dt$
where $\eta_i(0)$ is the noise at $t=0$ that kicks the $i$-th particle.
Now, starting from this `new' initial coordinates ${x_i}'$ the maximum
of the system has to subsequently stay below $L$ till the stopping time.
Finally, one must sum over all possible values of the new coordinates
${x_i}'$. Thus one must have
\begin{equation}
Q_N\left(x_1,x_2,\ldots, x_N\big|L\right)= \langle Q_N\left(x_1+\eta_1(0)dt, 
x_2+\eta_2(0)dt,\ldots ,x_N+\eta_N(0)dt\big|L\right)\rangle
\label{evol1}
\end{equation}
where the $\langle .\rangle$ denote the averages over the initial noises
$\eta_i(0)$. Expanding the right hand side in a Taylor series and using
(i) $\langle \eta_i(0)\rangle =0$ and (ii) $\langle 
\eta_i(0)\eta_i(0)\rangle= 1/dt$
(which follows from the delta correlator), one finds that $Q_N(\vec x\big|L)$
satisfies the Laplace's equation in the $N$-dimensional hypercube, $0\le x_i\le  L$,
\begin{equation}
\nabla^2 Q_N(x_1,x_2,\ldots x_N\big|L)=0.
\label{laplace1}
\end{equation}
The information about the maximum is captured in the boundary conditions.
For example, if $x_i=0$ for any $i$, $Q_N=1$ since if the $i$-th particle
starts at the origin, the process stops immediately ($t_s=0$) and hence
the maximum is necessarily (with probability $1$) less than $L$.
On the other hand, if $x_i=L$ for any $i$, one has $Q=0$. This follows
from the fact that if the $i$-th particle starts at $L$, it will
immediately cross the level $L$ and the probability that the maximum
will stay below $L$ till $t_s$ is necessarily zero. 

To summarize, $Q_N(\vec x\big|L)$ satisfies Laplace'e equation \eqref{laplace1}
in an $N$-dimensional hypercube $0\le x_i\le L$ with $Q=1$ for
any $x_i=0$ and $Q=0$ for any $x_i=L$. 
Thus, it reduces to an electrostatic problem where
one needs to find the potential $Q_N(\vec x\big|L)$ inside 
the hypercube $[0,L]^N$, whose $N$
faces touching the origin are held at a constant potential
$Q_N=1$ while the rest of the $N$ faces are earthed ($Q_N=0$).
We present the solutions explicitly for $N=1$ and $N=2$ in the next two
subsections.

\vskip 0.3cm 

\noindent {\bf Scaling:} Let us remark that since the only length scales are
the intial positions $\{x_i\}$ of the particles and the size
of the box $L$, it is evident that the exit 
probability $Q_N(\vec x\big|L)$
satisfies the scaling property
\begin{equation}
Q_N(\vec x\big|L) = 
Q_N\left(\frac{x_1}{L},\frac{x_2}{L},\ldots,\frac{x_N}{L}\right)=
Q_N(z_1,z_2,\ldots,z_N)
\label{scaling1}
\end{equation}
where the dimensionless scaled variables $0\le z_i=x_i/L\le 1$.

\vskip 0.3cm 

\noindent {\bf Special initial condition and a duality relation:} It is useful to 
consider a special initial condition where all the particles start
from the same initial positions: $x_i=x$ for all $1\le i \le N$.
In this case, the exit probabality is a function of 
a single scaled variable $0\le z=x/L\le 1$: $Q_N(x_1=x,x_2=x,\ldots, 
x_N=x\big|L)=q_N(x/L)$ where the scaling function $q_N(z)$ satisfies 
\begin{equation}
q_N(z)+q_N(1-z)=1.
\label{symm1}
\end{equation}
This duality relation states that the exit probability through the right 
boundary at $L$ starting from the initial position $L-x$ (of all 
the particles) is exactly identical to the exit probability through the left boundary $0$ starting from the initial positions $x$ (of all the particles). The duality relation 
\eqref{symm1} in particular dictates that 
\begin{equation}
q_N(z=1/2)=1/2
\label{symm2}
\end{equation}
for all $N$. The general solution of the Laplace's equation must satisfy 
Eq.~\eqref{symm1} which actually provides a useful check for the validity of
the solution.

\subsection{$N=1$}

For $N=1$, we have a second order ordinary differential equation  
\begin{equation}
\frac{d^2Q_1}{dx_1^2}=0 \quad {\rm with}\,\, Q_1(0\big|L)=1\,\, {\rm 
and}\,\, 
Q_1(L\big|L)=0,
\label{n1lap}
\end{equation}
whose solution is 
\begin{equation}
Q_1(x_1\big|L)= 1-\frac{x_1}{L}; \quad {\rm} 0\le x_1\le L
\label{n1lap1}
\end{equation}
Since $Q_1(x_1\big|L)=Q_1(L\big|x_1)= {\rm Prob}[m\le L\big|x_1]$, it 
follows, by 
differentiation that
the probability density of the maximum $m$ has a strict power law
form for all $m$
\begin{equation}
P_1(m\big|x_1)= \frac{x_1}{m^2} \quad {\rm for}\,\, m\ge x_1
\label{n1pdf}
\end{equation}
which is normalized to unity over $m\in [x_1,\infty]$ and all its 
integer moments diverge~\cite{KM1}. Thus the prefactor $B_1=1$
in \eqref{max1}.

Clearly for $N=1$, the exit probability $Q_1(x_1\big|L)$ is only a function 
of the 
scaled
variable $z=x_1/L$: $Q_1(x_1\big|L)=q_1(x_1/L)$ where the scaling function
$q_1(z)$ is simple
\begin{equation}
q_1(z)=1-z
\label{q1z}
\end{equation}
and evidently it satisfies the duality relation \eqref{symm1}.

\subsection{$N=2$}
\begin{figure}
\includegraphics[height=6.0cm,width=7.0cm,angle=0]{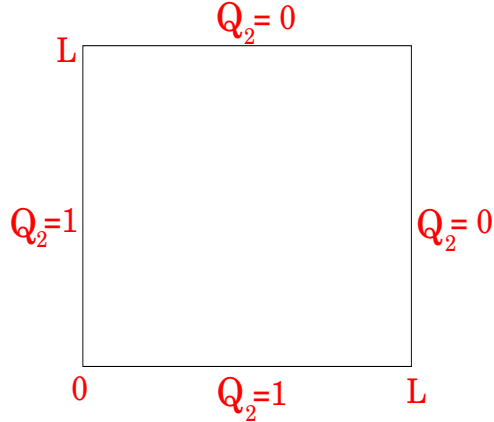}
\caption{\label{fig:square} Laplace's equation to be solved in
a square $(L\times L)$ with boundary conditions as shown.}
\end{figure}

The solution of the Laplace's equation becomes more involved 
in higher dimensions $N>1$. For $N=2$, one needs
to solve Laplace's equation in an $(L\times L)$ square
with boundary conditions shown in Fig. \ref{fig:square}.
The solution can be written down explictly for this case, although
the expression is rather cumbersome. Using separation
of variables one finds the appropriate solution~\cite{lebedev}
\begin{equation}
Q_2(x_1,x_2\big|L)= 1- \frac{x_1}{L} +\frac{2}{\pi}\, \sum_{n=1}^{\infty} 
\frac{(-1)^{n-1}}{n}\, 
\frac{\left[\sinh\left(n\pi\left(1-\frac{x_2}{L}\right)\right)
+(-1)^n \sinh\left(\frac{n\pi x_2}{L}\right)\right]}{\sinh(n\pi)}\, 
\sin\left(\frac{n\pi x_1}{L}\right).
\label{n2lap1}
\end{equation}

It is straightforward to verify that \eqref{n2lap1} is a solution of the Laplace's equation. Next we need to check that it satisfies the $4$ boundary conditions (see 
Fig. \ref{fig:square}). It is easy to check the two conditions:
(i) $Q_2(x_1=0,x_2\big|L)=1$ for all $x_2$ and (ii) $Q_2(x_1=L,x_2\big|L)=0$ 
for
all $x_2$. The other two conditions can also be verified. For
instance, putting $x_2=L$ in \eqref{n2lap1} we get
\begin{equation}
Q_2(x_1,L\big|L)= 1-\frac{x_1}{L} - \frac{2}{\pi}\sum_{n=1}^{\infty} 
\frac{1}{n}\,\sin\left(\frac{n\pi x_1}{L}\right).
\label{n2bc1}
\end{equation} 
Using the identity
\begin{equation}
\sum_{n=1}^{\infty} \frac{1}{n}\,\sin\left(\frac{n\pi 
x_1}{L}\right)=\frac{\pi}{2}\left(1-\frac{x_1}{L}\right)
\label{n2bc2}
\end{equation}
we verify that $Q_2(x_1,L\big|L)=0$. Similarly, putting
$x_2=0$ in \eqref{n2lap1} and using the identity
\begin{equation}
\sum_{n=1}^{\infty} \frac{(-1)^{n-1}}{n} \sin\left(\frac{n\pi 
x_1}{L}\right)=\frac{\pi\, x_1}{2L}
\label{n2bc3}
\end{equation}
one verifies the last boundary condition $Q_2(x_1,x_2=0\big|L)=1$.
Note that even though the solution $Q_2(x_1,x_2\big|L)$ in \eqref{n2lap1}
is not manifestly symmetric under the exchange of $x_1$ and $x_2$,
it is actually symmetric in $x_1$ and $x_2$
as it should be. Later in Section 4, we will derive an alternative 
expression via the path integral method which is manifestly symmetric
in $x_1$ and $x_2$.

Let us analyse the large $L$ behavior of \eqref{n2lap1} in the situation when 
$x_1$ and $x_2$ are kept finite. Since $Q_2(x_1,x_2\big|L)=Q_2(x_1/L,x_2/L)$ is 
a function of only the scaled variables $z_1=x_1/L$ and $z_2=x_2/L$, 
the $L\to \infty$ limit  
is equivalent to taking the limits $z_1\to 0$ and $z_2\to 0$.
Clearly $Q_2(z_1=0,z_2=0)=1$. It is easy to check that the first
derivatives $\partial_{z_1}Q$ and $\partial_{z_2}Q$ vanish
at the origin $(z_1=0,z_2=0)$. Similarly, the second
derivatives $\partial^2_{z_1}Q_2$ and $\partial^2_{z_2}Q_2$ also
vanish at the origin. So, in a Taylor expansion of $Q_2(z_1,z_2)$
the first nonzero term is the cross derivative, indicating
the following leading order behavior as $z_1\to 0$ and $z_2\to 0$
\begin{equation}
Q_2(z_1,z_2)= 1 - A_2 z_1 z_2 + \ldots
\label{n2taylor1}
\end{equation}
The amplitude $A_2$ is given by 
\begin{equation}
A_2= \frac{\partial^2 Q_2}{\partial z_1 \partial z_2}\big|_{z_1=0,z_2=0}= 
2\pi\,\left[\sum_{n=1}^{\infty} (-1)^n\, n\, \coth(n\pi)
-\sum_{n=1}^{\infty} \frac{n}{\sinh(n\pi)}\right]
\label{a21}
\end{equation}
where the last line follows from \eqref{n2lap1}. Fortunately, the 
sums can be carried out explicitly using some known
identities~\cite{Prudnikov} to yield 
\begin{equation}
A_2= \frac{1}{8\pi^2}\left[\Gamma(1/4)\right]^{4}=2.18844\dots
\label{a22}
\end{equation}
Hence for large $L$ 
\begin{equation}
{\rm Prob}\left[m\le L\big|x_1,x_2\right]=Q_2(x_1,x_2\big|L)\simeq 1- A_2 
\frac{x_1\,x_2}{L^2}.
\label{n2cum1}
\end{equation}
which leads to the announced power law
tail for the probability density of the maximum $m$
\begin{equation}
P_2(m\big|x_1,x_2)\simeq B_2 \frac{x_1\,x_2}{m^3}; \quad {\rm with}\,\, 
B_2=2A_2=
4.37688\ldots
\label{n2pdf1}
\end{equation}

To compare with the $N=1$ case, let us consider the special initial
condition where both particles start from the same initial position:
$x_1=x_2=x$. In this case, the exit probability 
$Q_2(x_1=x,x_2=x\big|L)=q_2(x/L)$ with scaling function $q_2(z)$ given by
\begin{equation}
q_2(z)= 1- z +\frac{2}{\pi}\, \sum_{n=1}^{\infty}
\frac{(-1)^{n-1}}{n}\,
\frac{\left[\sinh\left(n\pi\left(1-z\right)\right)
+(-1)^n \sinh\left(n\pi z\right)\right]}{\sinh(n\pi)}\,
\sin\left(n\pi z\right).
\label{q2z}
\end{equation}
One can verify that $q_2(z)$ satisfies the duality relation \eqref{symm1}.
Near $z=0$ and $z=1$, the scaling function has the asymptotics
\begin{equation}
q_2(z) \simeq
\begin{cases}
 1- A_2\, z^2 \quad\quad {\rm as}\,\, z\to 0, \\
A_2\,(1-z)^2 \quad\,\, {\rm as}\,\, z\to 1. 
\end{cases}
\label{q2zright}
\end{equation}
Comparing with the expression for the $N=1$ case, Eq.~\eqref{q1z}, one finds
that $q_2(z)$ has a much richer functional form. An analytic prediction for 
the function $q_2(z)$ is in excellent agreement with results
obtained from Monte Carlo simulations (Fig. \ref{fig:exitprob}). 

\section{Exact Solution for all $N$ by Path Counting Method}

The approach based on the Laplace equation (Sect.~\ref{Lap-sec}) is 
difficult to extend for $N>2$. The technical
problem is to find the exact solution of the Laplace equation
$\nabla^2 Q_N(z_1,z_2,\ldots, z_N)=0$ in the $N$-dimensional hypercube
of the scaled variables $0\le z_i=x_i/L\le 1$, with the prescribed
boundary conditions that $Q_N=1$ for any $z_i=0$ and $Q_N=0$ for
any $z_i=1$. For $N>2$, it is not easy to find an explicit solution
to this problem. In this section, we use an alternative path counting
method that is physically more explicit and, in addition, allows
us to write down the exact solution $Q_N(z_1,z_2,\ldots, z_N)$ for all
$N$. This method thus provides an alternative way to solving the Laplace's
equation in a hypercube with the prescribed boundary conditions.

To set up the path counting method for general $N$, we need two basic ingredients
from the single particle problem in a box $[0,L]$ with absorbing boundary 
conditions at the two boundaries $0$ and $L$:
\vskip 0.3cm

\noindent (i) the survival probability
$S(x_0,t,L)$ which counts the probability that a single particle,
starting at $x_0$ at time $0$, remains inside the box $[0,L]$ during the
time interval $(0,t)$.

\vskip 0.2cm

\noindent (ii) the first passage probability density $F(x_0,t,L)$ that
denotes the probability density that the particle, starting initially at 
$x_0$ ($0\le x_0\le L$), exits 
for the first time the box through  
the boundary at $0$ (and not through the other boundary at $L$) at time $t$. 

\vskip 0.2cm

One can write the survival probability $S(x_0,t,L)=\int_0^L G(x,x_0,t,L)\, dx$
where $G(x,x_0,t,L)$ denotes the Green's function counting the probability
density that the particle reaches $x$ at time $t$, starting from $x_0$ at 
time $0$, while staying inside the box $[0,L]$ during time $t$. This
Green's function can be computed (a) either by solving the diffusion 
equation $\partial_t G=D\partial_x^2 G$ (the diffusion constant is $D=1/2$ for our choice of the noise term) with absorbing boundary condition $G(x=0,x_0,t,L)=G(x=L,x_0,t,L)=0$ and the initial condition $G(x,x_0,0,L)= \delta(x-x_0)$,
or (b) by the path integral method. The resulting Green's function admits the 
following representation 
\begin{equation}  
G(x,x_0,t,L)= \frac{2}{L}\sum_{n=1}^{\infty} \sin\left(\frac{n\pi x}{L}\right)\,
\sin\left(\frac{n\pi x_0}{L}\right)\, e^{-\frac{n^2\pi^2}{2L^2}\,t}.
\label{green1}
\end{equation}
The survival probability, after integrating over the final position $x$, is 
\begin{equation}
S(x_0,t,L)= \frac{2}{\pi}\sum_{n=1}^{\infty} \frac{\left[1-(-1)^n\right]}{n}\,
\sin\left(\frac{n\pi x_0}{L}\right)\,  e^{-\frac{n^2\pi^2}{2L^2}\,t}.
\label{surviv1}
\end{equation}

To calculate the first passage probability density $F(x_0,t)$ through $0$ at time 
$t$, we note that this just counts the flux of particles going out of the
box through $0$ at time $t$. The flux through a point $x$ in the positive
direction is the probability current $-D \partial_x G$ (with $D=1/2$). Hence, the 
first-passage probability through the origin that counts the flux
through the origin in the negative direction is simply,
$F(x_0,t)=D \partial_x G\big|_{x=0}$. Using $G$ from \eqref{green1}, we get
\begin{equation}
F(x_0,t)= \frac{\pi}{L^2} \sum_{n=1}^{\infty} n\, \sin\left(\frac{n\pi 
x_0}{L}\right)\,  e^{-\frac{n^2\pi^2}{2L^2}\,t}.
\label{fp1}
\end{equation}  

Armed with these two ingredients from the single particle problem, we are now 
ready to compute the exit probability $Q_N(\vec x\big|L)$ for the 
$N$-particle
problem. Consider first the event shown in Fig. {\ref{fig:mult} where
one of the particles (say the $i$-th one) exits the box for the first time
between time $t_s$ and $t_s+dt_s$ while the $N-1$ other particles stay
inside $[0,L]$ till this time $t_s$. Clearly, the probability
for this event, using independence of walkers, is given by
the product $F(x_i,t_s)\,\prod_{j\ne i} S(x_j,t_s)\, dt_s$.
Now, the particle that hits $0$ (whose label is $i$) can, in fact,
be any one of the $N$ particles. Hence we have to sum over the hitting
index $i$ from $1$ to $N$. In addition, this event can occur
at any time $t_s$, so we need to integrate
over $t_s$. This path counting method then gives the following net contribution 
to the exit probability
\begin{equation}
Q_N(\vec x\big|L)= \sum_{i=1}^N \int_0^{\infty} F(x_i,t_s)\,\prod_{j\ne i} 
S(x_j,t_s)\, dt_s.
\label{exitp1}
\end{equation}
Substituting the results for $S(x,t)$ and $F(x,t)$ respectively from
Eqs. (\ref{surviv1}) and (\ref{fp1}), integrating over $t_s$ and
setting $z_i=x_i/L$ 
gives our main exact result, valid for all $N$,
\begin{equation}
Q_N(\vec x\big|L)= Q_N(z_1,z_2,\ldots,z_N)=\left(\frac{2}{\pi}\right)^N\, 
\sum_{\{n_i\}} 
\frac{a(n_1,n_2,\ldots,n_N)}{n_1^2+n_2^2+\ldots+ n_N^2}\, \prod_{k=1}^N 
\frac{\sin(n_k \pi z_k)}{n_k}
\label{exitp2}
\end{equation}
where each index $n_i=1,2,\ldots$ for all $1\le i\le N$ and
\begin{equation}
a(n_1,n_2,\ldots,a_N)= \sum_{i=1}^N n_i^2 \prod_{j\ne i} 
\left[1-(-1)^{n_j}\right].
\label{afunction}
\end{equation}
Equation \eqref{exitp2} is an exact solution of
the Laplace's equation in the $N$-dimensional hybercube $0\le z_i\le 1$ with 
the prescribed boundary conditions which has a virtue of being manifestly  symmetric with respect to the interchange of the coordinates $z_i$'s. 

To extract the behavior in the limit of large $L$, with $x_i$'s fixed,
we need to take the limit $z_i\to 0$ in Eq. (\ref{exitp2}). 
This limiting behavior reads
\begin{equation}
Q_N(z_1,z_2,\ldots, z_N)= 1- A_N z_1\,z_2\ldots z_N + O(z_i^2)
\label{exitp3}
\end{equation}
The coefficient $A_N$ can be computed by taking the derivatives of 
Eq.~\eqref{exitp2} with respect to each $z_i$ and then putting $z_i=0$ for all $i$.
This gives
\begin{equation}
A_N= -N\, 2^N\, \lim_{z_i\to 0}\, 
\sum_{\{n_i\}}\frac{n_1^2\cos(n_1\pi z_1)}{n_1^2+n_2^2+\ldots+n_N^2}
\prod_{j=2}^N \left[1-(-1)^{n_j}\right]\cos(n_j \pi z_j).
\label{coeff1}
\end{equation}

Noting that $Q_N(\vec x\big|L)=Q_N(z_1,z_2,\ldots,z_N)$ is precisely the
cumulative distribution of the maximum $m$, one thus gets
\begin{equation}
{\rm Prob}[m\le L\big|\vec x]\simeq 1- A_N \frac{x_1x_2\ldots x_N}{L^N}\quad 
{\rm as}\, 
L\to \infty.
\label{maxcum1}
\end{equation}
Differentiating with respect to $L$ and putting $L=m$ gives the exact
power-law tail of the pdf of the maximum $m$ in Eq. (\ref{max1})
with the prefactor $B_N$ is given by the formal sum
\begin{equation}
B_N= N A_N = -N^2\, 2^N\, \lim_{z_i\to 0}\,
\sum_{\{n_i\}}\frac{n_1^2\cos(n_1\pi z_1)}{n_1^2+n_2^2+\ldots+n_N^2}
\prod_{j=2}^N \left[1-(-1)^{n_j}\right]\cos(n_j \pi z_j)
\label{prefact1}
\end{equation}
where each index $n_i$ runs over all positive integers.
  
The formal sum in Eq. (\ref{prefact1}) can be explicitly evaluated for
$N=1$ and $N=2$. For $N=1$
\begin{equation}
B_1= -2  \lim_{z_1\to 0}\, \sum_{n_1=1,2,\ldots} \cos(n_1 \pi z_1)=1
\label{BN1}
\end{equation}
in agreement with already known results. For $N=2$,  equation \eqref{prefact1} gives
\begin{equation}
B_2= - 16 \lim_{z_i\to 0}\, \sum_{n_1,n_2} \frac{n_1^2\, 
\cos(n_1\pi z_1)}{n_1^2+n_2^2}\, \left[1-(-1)^{n_2}\right]\cos(n_2\pi z_2)
=\frac{1}{4\pi^2}\left[\Gamma\left(\frac{1}{4}\right)\right]^{4}=4.37688\ldots
\label{BN2}
\end{equation}
In Appendix A, we show how to compute the above sum explicitly.

For $N>2$, we haven't deduced explicit expressions for the sum 
in \eqref{prefact1}. However,
one can reduce it to a simpler form where the sum is
rapidly convergent and can then be evaluated by Mathematica.
As an example, for $N=3$, 
\begin{equation}
B_3= -6 + 
\frac{9}{8\pi^2}\Gamma^4(1/4)+72\pi\sum_{n_1,n_2}
\frac{n_1^2+n_2^2\, (-1)^{n_2}}{\sqrt{n_1^2+n_2^2}\,
\sinh\left(\pi\sqrt{n_1^2+n_2^2}\right)}=15.3369\dots
\label{BN3}
\end{equation}
In the limit of large $N$, one can evaluate the formal sum (see Appendix B)
to obtain the limiting behavior
\begin{equation}
B_N \simeq N\left[\frac{4}{\pi}\ln(N)\right]^{N/2}
\label{ampN2} 
\end{equation} 
in perfect agreement with the heuristic result in Eq. (\ref{ampNheu}).

\vskip 0.3cm

\noindent {\bf Special initial condition:} Finally, let us consider the special 
initial condition when all the particles 
start from the same point: $x_1=x_2=\ldots=x$ where $0\le x\le L$.
In this case, the exit probability in Eq. (\ref{exitp2}) reduces
to a function of one scaled variable $z=x/L$:
$Q_N(x,x,\ldots,x \big| 
L)=q_N(z)$ given by 
\begin{equation}
q_N(z)= N\, \left(\frac{2}{\pi}\right)^N \, \sum_{\{n_i\}} \frac{n_1 \sin(n_1\pi 
z)}{(n_1^2+n_2^2+\ldots+n_N^2)} \prod_{j\ne 1} \left[1-(-1)^{n_j}\right]\, 
\frac{\sin(n_j \pi z)}{n_j}.
\label{qNz1}
\end{equation}
One can check that $q_N(z)$ satisfies the duality relation $q_N(z)+q_N(1-z)=1$
and thus $q_N(1/2)=1/2$ for all $N$.

Performing this multiple sum directly by Mathematica is difficult as it 
converges slowly. To circumvent this problem,
we first perform the sum over $n_1$ in Eq. (\ref{qNz1}) using the following 
identity~\cite{Prudnikov}
\begin{equation}
\sum_{k=1}^{\infty} \frac{k \sin(k\pi x)}{k^2+a^2}= \frac{\pi}{2}\, \frac{\sinh 
\left(\pi(x-a)\right)}{\sinh(\pi a)}.
\label{iden-sin}
\end{equation}
This gives
\begin{equation}
q_N(z)= N\,  \left(\frac{4}{\pi}\right)^{N-1} \, \sum_{n_2,n_3,\ldots,n_N\to 
{\rm odd}} 
\frac{\sinh\left[\pi(1-z)
\sqrt{n_2^2+n_3^2+\ldots+n_N^2}
\right]}{\sinh\left[\pi\sqrt{n_2^2+n_3^2+
\ldots+n_N^2}\right]}\, \prod_{j=2}^N \frac{\sin(n_j \pi z)}{n_j}.
\label{qNz2}
\end{equation} 
The multiple sum in Eq. (\ref{qNz2}) is now rapidly convergent and can be
easily evaluated by Mathematica. In Fig. \ref{fig:mult} we plot this function
$q_N(z)$ for $N=2$ and $N=3$. For $N=2$, it of course coincides
with the earlier expression \eqref{q2z} obtained via the Laplace's method
in the previous section.
\vskip 0.3cm

\noindent{ \bf Average Maximum:} For fixed identical initial position of 
all particles $x_i=x$,
another interesting question is how does the average maximum (till
the stopping time $t_s$) depend on $N$? For $N=1$ the average
is infinite but for all $N>1$ it is finite. However, does the average maximum
for $N>1$  
increase or decrease as the number of walkers $N$ increases?
The answer to this question is not intuitively obvious.
However, knowing the function $q_N(z)$, one can
compute the average maximum in the following way.
We have, ${\rm Prob}[m\le L|x]= q_N(x/L)$ where
$q_N(z)$ is given in Eq. (\ref{qNz2}). Therefore the probability density 
function of $m$ reads 
\begin{equation}
P_N(m|x)= -\frac{x}{m^2} q_N'\left(\frac{x}{m}\right);\quad\ 
m\ge x
\label{pdf1}
\end{equation}
where $q_N'(z)= dq_N(z)/dz$.
The first moment is then given by, for all $x\ge 0$ and $N>1$
\begin{equation}
\langle m\rangle = -\int_{x}^{\infty} 
\frac{x}{m}\,q_N'\left(\frac{x}{m}\right)\, 
dm= C_N\, x
\label{avg1}
\end{equation}
with prefactor
\begin{equation}
C_N= -\int_0^1 \frac{q_N'(z)}{z}\, dz. 
\label{cn1}
\end{equation}
Thus the average maximum, for $N>1$,  is proportional to $x$ for all $x$ and 
the
proportionality constant $C_N$ is given by Eq. (\ref{cn1}). We were
unable to carry out the integral in Eq. (\ref{cn1}) in closed form.
However, it is clear that as $N\to \infty$, using $q_N(z)= \theta(1/2-z)$,
one gets $C_N\to 2$. On the other hand $C_N$ diverges as $N\to 1$. Thus,
$C_N$ decreases when $N$ increases. These
results are supported by Monte Carlo simulations (Fig.~\ref{fig:maxavg}).

\section{Numerical Simulations}

Our analytical result for the exit probability 
through the origin $q_N(z)$, when 
all particles start at the same scaled position 
$0\le z=x/L\le 1$, is tested by Monte 
Carlo simulations. For a fixed box of size $L$, 
the exit probility for the $N$-particle problem can be efficiently computed 
using a method proposed in Ref.~\cite{MRZ1}.
Naively, to compute $q_N(z)$, one would first fix the starting point 
$0\le z\le 1$ of all the Brownian motions and then generate 
different realizations of the process and compute
the fraction of realizations where the first exit happens
through the origin. One would then repeat the procedure
for each value of $0\le z\le 1$ (with an appropriate bin size) in order 
to compute the
full function $q_N(z)$ over the range $z\in [0,1]$. 
Instead of repeating the simulation for each starting point $z$,
it turns out to be more efficient to follow a different algorithm 
described briefly below.

For illustration, we take the example of just one Brownian motion ($N=1$).
The method is easily generalized for all $N$. We start the Brownian
motion at the origin, let it evolve in time and  
record the maximal 
($x_{\text{max}}$) and the minimal ($x_{\text{min}}$) position reached by the 
walker up to time $t$. The process 
is halted when 
$x_{\text{max}}- x_{\text{min}} \geq L$ for the first time 
(see Fig. \ref{fig:algo}). 
Keeping $x_{\text{min}}$ and $x_{\text{max}}$ fixed, we  
now horizontally slide the whole configuration thereby changing
the starting point (see the second and the third panel of Fig. 
\ref{fig:algo}). Measuring all distances with respect to $x_{\rm min}$,
it is then clear that this configuration contributes $1$ to $q_1(z)$ for
$z\in [0,1-x_{\text{max}}/L]$ and $0$ for $z\in [1-x_{\text{max}}/L,1]$.
So, for this configuration, we just record the number $y=1-x_{\text{max}}/L$
and the fact that it contributes $\theta[y-z]$ to $q_1(z)$. 
We then repeat this procedure for another configuration starting at $0$,
do the sliding and record the value of $y$ and the associated
$\theta(y-z)$. We repeat this, say, for $N_s$ number of samples. 
To sum all the contributions, we first sort
the values of $y$ associated with the configurations in increasing
order. Let $\{{\tilde y}_1, {\tilde y}_2, \ldots, {\tilde y}_{N_s}\}$
denote the sorted values. Then, we assign $q_1({\tilde y}_1)=1$,
$q_1({\tilde y}_2)=1-1/N_s$ and in general $q_1({\tilde y}_i)=1-(i-1)/N_s$.
This generates the full curve
$q_1(z)$.

Thus this algorithm has two advantages: (i) it does not require
to repeat the simulations for each value of $z$ and (ii) it does not
require any specific choice of bin sizes.
This clearly
makes the simulation much faster.
The results of our simulations for $N=2$, $N=3$ and $N=6$ are plotted in Fig. 
(\ref{fig:exitprob}). The agreement with the analytical prediction is 
excellent. For all $N$, $q_N(z)$ satisfies the duality relation: 
$q_N(z)+q_N(1-z)=1$.
\begin{figure} 
\includegraphics[width=14.0cm]{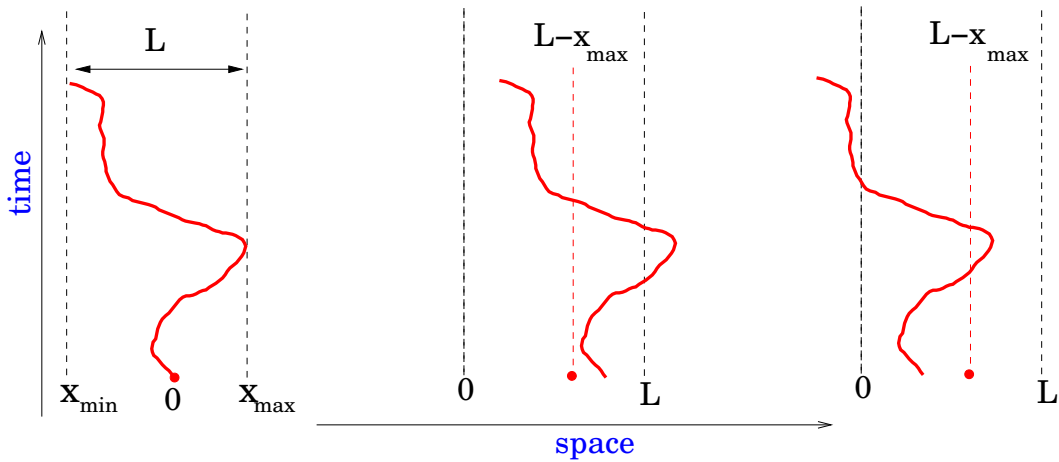}
\caption{\label{fig:algo} The illustration of the algorithm
to compute the exit probability $q_N(z)$ for a single ($N=1$)
Brownian motion.}
\end{figure}

In Ref.~\cite{MRZ1} we had studied the exit probability
from a box $[0,L]$ through the upper boundary $L$ (rather than
the lower boundary $0$ considered here) for
a single particle whose motion is described by a generic self-affine 
stochastic process. This probability was called the
`hitting probability' in \cite{MRZ1}. 
In the present paper, we are considering the complimentary
event of exiting the box $[0,L]$ through the lower boundary $0$.
In the notation of the
present paper, the hitting probability (of the boundary $L$)
would correspond to $1-q_N(z)$~\cite{notation}.
For a generic self-affine process, it was shown in Ref. \cite{MRZ1} 
that the exit probability should have a power law behavior close to the origin 
$1-q(z)\sim z^{\phi}$ with $\phi=\theta/H$ as mentioned in the
introduction. In addition, it was observed in ref. \cite{MRZ1} that
for many processes (but not all), once we know the exponent $\phi$, the full 
function  
$q(z)$ over the range $z\in [0,1]$ is described by a universal
one-parameter (parametrized by $\phi$) form~\cite{MRZ1}
\begin{equation}
q_\phi(z)=1-I_z(\phi,\phi)=
1-\frac{\Gamma(2 \phi)}{\Gamma(\phi)^2} \int_0^z [u(1-u)]^{\phi-1} d u .
\label{univ1}
\end{equation}
The function $I_z(\phi,\phi)$ is the incomplete regularized beta function. In 
our present problem, we have seen 
in Eq. (\ref{exitp3}) that for small $z$, $q_N(z)\sim 1-A_N z^N$
indicating $\phi=N$.
It is then natural to investigate if our result for the full function $q_N(z)$
can be re-expressed as the universal functional form in Eq. (\ref{univ1})
with $\phi=N$. Interestingly, the answer is no, as it is clearly 
shown in Fig \ref{fig:hypergeo}, thus providing us with a
counterexample. 

We have also computed the prefactor $B_N$ numerically from
the Monte Carlo simulations up to $N=7$.
The results are shown in Fig.  \ref{fig:BN} by squares.
For $N=1$, $N=2$ and $N=3$, they agree with our exact analytical
predictions $B_1=1$, $B_2=4.37688\dots$ and $B_3=15.3369\dots$
as discussed in Section IV. Had the $q_N(z)$ be described by
the universal function in Eq. (\ref{univ1}) with $\phi=N$, 
one would get from the small $z$ expansion in Eq. (\ref{univ1}),
$q_N(z)\to 1- A_N z^N$ with $A_N= \Gamma(2N)/{N\Gamma^2(N)}$.
This would predict the prefactor $B_N= NA_N= \Gamma(2N)/{\Gamma^2(N)}$.
In  Fig.  \ref{fig:BN}, this prediction from the universal curve
is shown by the dashed line. Clearly, it does not match the 
simulation results, confirming once more that $q_N(z)$ is
not described by the universal function in  Eq. (\ref{univ1}).  
We have also plotted the exact asymptotic prediction of $B_N$
in Eq. (\ref{ampN2}) as a solid line for comparison.
While it is difficult to extract the small $z$ behavior of $q_N(z)$ 
and hence $B_N$ for larger values of $N$, we 
notice that the asymptotic large $N$ behavior is already approached
for $N=7$.

Finally we have also computed the average maximum (till
the stopping time $t_s$) of $N$ Brownian motions starting from 
the same initial positions $x>0$. We verified that for all $N>1$, 
the average maximum exists, and it is proportional to $x$, as
predicted analytically in Eq. (\ref{avg1}). In Fig. \ref{fig:maxavg} we plot the 
amplitude $C_N$, given analytically in Eq. (\ref{cn1}),  as a function of $N$. As 
predicted, $C_N$ approaches $2$  for large $N$.

\begin{figure}
\includegraphics{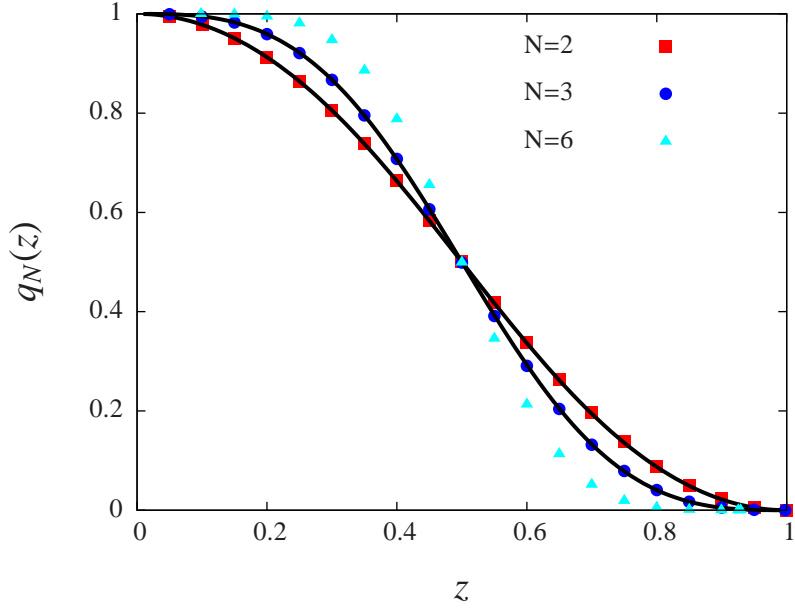}
\caption{\label{fig:exitprob} The exit probability $q_N(z)$
through the origin
obtained from Monte Carlo simulations for $N=2$ (red squares),
$N=3$ (deep blue circles) and $N=8$ (blue triangles).
The numerical data are compared 
to analytical results (solid lines) for   
$N=2$ 
and $N=3$ (obtained from Eq. (\ref{qNz2}) respectively for $N=2$
and $N=3$). }
\end{figure}

\begin{figure}
\includegraphics{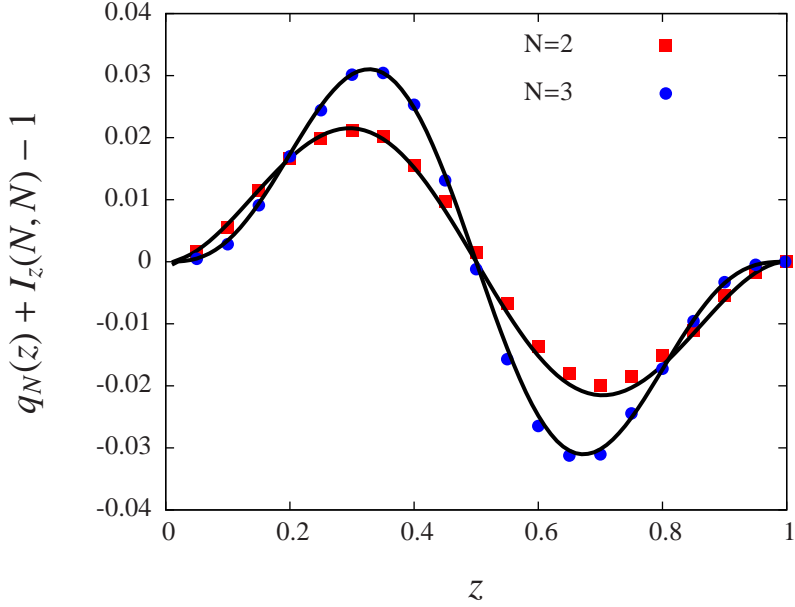}
\caption{\label{fig:hypergeo} 
Difference between the exit probability for $N=2$ and $N=3$  and the
universal function in Eq. (\ref{univ1}) with 
$\phi=N$. Solid lines are the
analytical results from Eq. (\ref{qNz2}) with $N=2$ and $N=3$. Symbols are 
the numerical results for
$10^5$ realizations of $N=2$ (squares) and $N=3$ (circles) Brownian motions 
starting from the same initial position in
a box of size $L=1000$.}
\end{figure}

%[height=8.0cm,width=8.0cm,angle=0]
\begin{figure}
\includegraphics{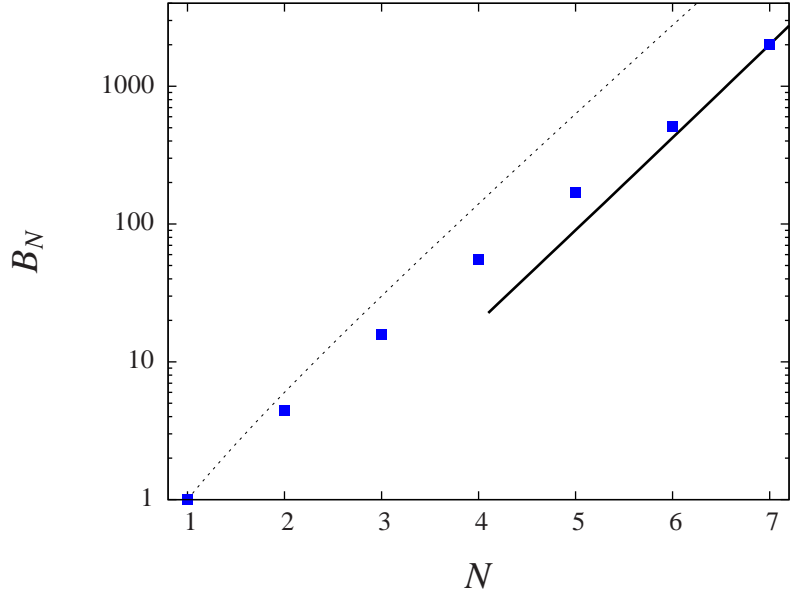}
\caption{\label{fig:BN} 
Dashed line: $B_N=\Gamma(2N)/{\Gamma^2(N)}$ from the universal
function in Eq. (\ref{univ1}). 
Solid line: aymptotic behavior for large N given in Eq.(\ref{ampN2}) 
and squares: 
direct simulations of
$N$ Brownian motions starting from the same position in a box of size
$L=1000$. Averages are performed over $10^6$ samples.}

\end{figure}
%[height=8.0cm,width=8.0cm,angle=0]
\begin{figure}
\includegraphics{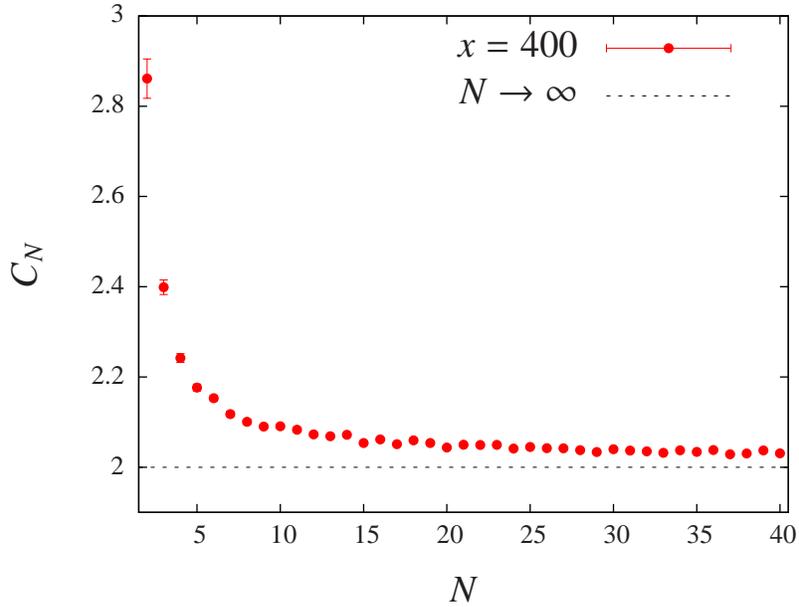}
\caption{\label{fig:maxavg}
Monte Carlo results for the average maximum of $N$ Brownian 
motions till the first passage time through the origin, all 
starting at the same initial position $x>0$. 
The coefficient  
$C_N= \langle m\rangle/x$, see Eq.~\eqref{cn1}, is plotted vs. $N$. Averages are 
performed over $10^3$ realizations.}
\end{figure}

\section{Conclusion}

To summarize, we have presented an exact solution for the probability
distribution of the maximum $m$ of a set of $N$ independent Brownian
motions starting at the initial positions ${\vec x}\equiv \{x_1,x_2,\ldots, 
x_N\}$ on the positive half-axis and the process terminating when any one
of the walkers crosses the origin. We have shown that for large $m$,
the pdf of $m$ decays as a power law, $P_N(m|\vec x)\simeq B_N (x_1x_2\ldots 
x_N)/m^{N+1}$ where the prefactor $B_N$ has an interesting $N$ dependence.
For a fixed $N>1$, integer moments of $m$ up to order $(N-1)$ are finite,  
while all higher integer moments are infinite. The cumulative distribution
of this maximum also provides an exact solution to the first-exit 
probability through the origin (rather than through $L$) of $N$ walkers from 
a box $[0,L]$. Incidentally, our path counting method  also provides an exact solution to the $N$-dimensional Laplace's equation $\nabla^2 Q_N=0$ in a 
hypercube $[0,L]^N$ with the boundary conditions $Q_N=1$ on any face of the hypercube passing through the origin and $Q_N=0$ on the rest of the faces.
Monte Carlo simulations confirm our analytical results.

This work raises some interesting open questions. We have focused
only on the maximum $m$ till the stopping time $t_s$ of $N$
independent walkers.  Another interesting observable is not just the actual
value of the maximum, but the time $t_m$ at which this maximum occurs
before the stopping time $t_s$. This random variable has recently been studied in
a number of contexts. For a stochastic process over a fixed time interval $[0,t]$,  
the distribution of the time $t_m$ has been computed 
for a variety of Brownian paths, such as a free Brownian motion, Brownian 
bridges, Brownian excursions and Brownian meanders, using path
integral method~\cite{MRKY} and also by an alternative functional 
renormalization group method~\cite{SL}. The distribution of $t_m$ was also computed exactly  
for the random acceleration process which is a non-Markov process~\cite{MRZ2}.
It has also been computed both
for independent Brownian walkers~\cite{RMC,chull} and very recently for vicious  
walkers~\cite{Rambeau-Schehr}. 
On the other hand, when the process stops at a random stopping time 
$t_s$ where for instance $t_s$ is the first time a walker hits the origin, 
the distribution of $t_m$ has been computed for a single Brownian
motion $(N=1)$ where it is already nontrivial~\cite{RM}. It would be
interesting to extend the results of Ref.~\cite{RM} to the case of
$N>1$ independent Brownian motions.

Another interesting challenging problem would be to compute the distribution 
of $m$ as well as that of $t_m$ for a set of vicious walkers till
the stopping time $t_s$ when the walker closest to the origin 
crosses the origin for the first time. 

\acknowledgments{PLK thanks Laboratoire de Physique Th\'{e}orique et 
Mod\`{e}les Statistiques for hospitality 
and the NSF grant CCF-0829541 for support. AR acknowledges
support by ANR grant 09-BLAN-0097-02.
We thank A. Schweiger for useful discussions.}

\appendix

\section{Evaluation of $B_2$}

To evaluate the sum in Eq. (\ref{BN2}) we first 
split the sum into two parts: $B_2= I_1+I_2$ with
\begin{eqnarray}
I_1 &= & -16 \lim_{z_1, z_2\to 0} \sum_{n_1,n_2} 
\frac{n_1^2}{n_1^2+n_2^2}\,\cos(n_1 \pi z_1)\cos(n_2 \pi z_2) 
\label{I1} \\
I_2 &=& 16 \lim_{z_1, z_2\to 0}  \sum_{n_1,n_2} 
\frac{(-1)^{n_2}\, n_1^2}{n_1^2+n_2^2}\,\cos(n_1 \pi z_1)\cos(n_2 \pi z_2)
\label{I2}
\end{eqnarray}
where all the sums run over positive integers.

Let us first evaluate $I_1$. Due to the symmetry we can re-write $I_1$ as
\begin{equation}
I_1= -8  \lim_{z_1, z_2\to 0} \sum_{n_1,n_2} 
\frac{n_1^2+n_2^2}{n_1^2+n_2^2}\,\cos(n_1 \pi z_1)\cos(n_2 \pi z_2)
= -8  \lim_{z_1, z_2\to 0}\sum_{n_1,n_2}\,\cos(n_1 \pi z_1)\cos(n_2 \pi z_2)
=-2
\label{I11}
\end{equation}
In the last step we have used the identity, $\sum_{n\geq 1} \cos(n\pi z)=-1/2$.

To evaluate $I_2$ we use another standard 
identity~\cite{Prudnikov}
\begin{equation}
\sum_{k=1}^{\infty} \frac{(-1)^{k} \cos(k\pi z)}{k^2+a^2}= 
\frac{\pi}{2a}\,\frac{\cosh(a\pi z)}{\sinh(a\pi)}-\frac{1}{2 a^2}.
\label{piden1}
\end{equation}
We now sum over $n_2$ in Eq. (\ref{I2}) using above identity. This gives
\begin{equation}
I_2=-8 \lim_{z_1\to 0} \sum_{n_1=1}^{\infty} \cos(n_1\pi z_1)+ 8\pi 
\lim_{z_1, z_2\to 0} \sum_{n_1=1}^{\infty} 
\frac{n_1}{\sinh(n_1\pi)}\, \cos(n_1\pi 
z_1)\cos(n_1 \pi z_2) = 4+ 8\pi \sum_{n_1=1}^{\infty}
\frac{n_1}{\sinh(n_1\pi)}.
\label{I21}
\end{equation}
The remaining sum in \eqref{I21} can be explicitly 
evaluated using the identity~\cite{Prudnikov}
\begin{equation}
\sum_{n=1}^{\infty} \frac{n}{\sinh(n\pi)}= \frac{1}{32 
\pi^3}\,\Gamma^{4}\left(\frac{1}{4}\right)-\frac{1}{4\pi}.
\label{piden2}
\end{equation}
Adding $I_1$ and $I_2$ we arrive at the announced expression \eqref{amp2}
for $B_2$. 

\section{Evaluation of $B_N$ for large $N$}

Here we show that $B_N$ in Eq. (\ref{prefact1}) behaves, to leading
order for large $N$, as
\begin{equation}
B_N\simeq N\left[\frac{4}{\pi}\ln(N)\right]^{N/2}.
\label{ampN3}
\end{equation}

First we rewrite the denominator in Eq. (\ref{prefact1}) using
the integral representation
\begin{equation}
\frac{1}{n_1^2+n_2^2+\ldots+n_N^2}= \int_0^{\infty} dt\, 
e^{-t\,(n_1^2+n_2^2+\ldots+ 
n_N^2)}.
\label{intrep1}
\end{equation}
Using this representation, one can then decouple the sums over
different indices $n_i$ in Eq. (\ref{prefact1}) giving
\begin{equation}
B_N= -N^2\,2^N\, \lim_{z_i\to 0}\, \int_0^{\infty} dt\, \sum_{n_1=1}^{\infty} 
n_1^2 \cos(n_1\pi 
z_1)\, 
e^{-n_1^2 t} \prod_{j\ne 1} \sum_{n_j=1}^{\infty} \left[1-(-1)^{n_j}\right]\, 
\cos(n_j \pi z_j)\, e^{-t n_j^2}  
\label{bn1}
\end{equation}
If we now take the limits $z_j\to 0$ inside the sums, each of the sums
is convergent. However, it is easy to check that the integrand, as
a function of $t$, diverges as $\sim t^{1+N/2}$ as $t\to 0$. Thus
the integral is longer convergent. To circumvent this difficulty,
one can use a standard regularization scheme used often in evaluating
the Madelung constant in the context of lattice sums in 
crystals~\cite{BBT} whereby we introduce a parameter $s$
and rewrite Eq. (\ref{bn1}) as, upon taking the limits $z_j\to 0$
inside the sums,
\begin{equation}
B_N(s) = -\frac{N^2\, 2^N}{\Gamma(s)}\, \int_0^{\infty} dt\, t^{s-1}\, 
\sum_{n_1=1}^{\infty}n_1^2 e^{-n_1^2\, t}
\, \left[\sum_{n_j=1}^{\infty} \left[1-(-1)^{n_j}\right]\, 
e^{-n_j^2\, t}\right]^{N-1}.
\label{bn2}
\end{equation}
Note that this integral is covergent for all $s>(1+N/2)$. The idea
is to first evaluate $B_N(s)$ for large $N$ with $s>(1+N/2)$ and then
analytically continue this result to $s\to 1$ to evaluate $B_N= B_N(s\to 1)$.

Let us next define the function
\begin{equation}
g(t) = 4\sqrt{\frac{t}{\pi}}\, \sum_{k=0}^{\infty} e^{-(2k+1)^2 t}.
\label{gt1}
\end{equation}
Up to the factor $\sqrt{t/\pi}$, this function $g(t)$ can be expressed
in terms of standard Jacobi theta functions~\cite{GR}. Then one can
rewrite Eq. (\ref{bn2}) as
\begin{equation}
B_N(s)= -\frac{2N^2 (\sqrt{\pi})^{N-1}}{\Gamma(s)}\,\int_0^{\infty} dt\, 
t^{s-(N+1)/2}\, \left[g(t)\right]^{N-1} \left[\sum_{n=1}^{\infty} n^2 
e^{-n^2\,t}\right].
\label{bn3}
\end{equation}
To evaluate this integral for large $N$, we need to know how
the function $[g(t)]^{N-1}$ behaves for large $N$. 

Let us first focus on the function $g(t)$ in Eq. (\ref{gt1}). Clearly, for 
large $t$, the dominant contribution comes from the $k=0$ term in the sum
and hence $g(t)\sim \sqrt{t} e^{-t}$ as $t\to \infty$. In contrast, the
opposite limit $t\to 0$ is more tricky. To derive its behavior as $t\to 0$,
we first use the following Jacobi identity~\cite{GR}
\begin{equation}
1+ 2\sum_{k=1}^{\infty} (-1)^k e^{-k^2\,z}= 2\sqrt{\frac{\pi}{z}}\, 
\sum_{k=0}^{\infty} e^{-(2k+1)^2 \pi^2/{4z}}
\label{jacobi1}
\end{equation}
to rewrite the function $g(t)$ (upon identifying $t=\pi^2/{4z}$) as
\begin{equation}
g(t) = 1+2\sum_{k=1}^{\infty} (-1)^k\, e^{-\pi^2 k^2/{4t}}.
\label{gt2}
\end{equation}
This representation of $g(t)$ is more amenable to the asymptotic analysis
in the $t\to 0$ limit. We obtain from Eq. (\ref{gt2}) the leading behavior
of $g(t)$ as $t\to 0$
\begin{equation}
g(t) \simeq 1- 2\, e^{-\pi^2/{4t} } +\ldots
\label{gt3}
\end{equation}
Thus $g(t)$ has an essential singularity at $t=0$ and it approaches
to $1$ as $t\to 0$ in an extremely flat way. Thus the function $g(t)$
starts at $g(0)=1$, stays flat for a while and then decreases exponentially
fast to $0$ as $t$ increases (see Fig. (\ref{fig:gt})). 
\begin{figure}
\includegraphics[height=8.0cm,width=8.0cm,angle=0]{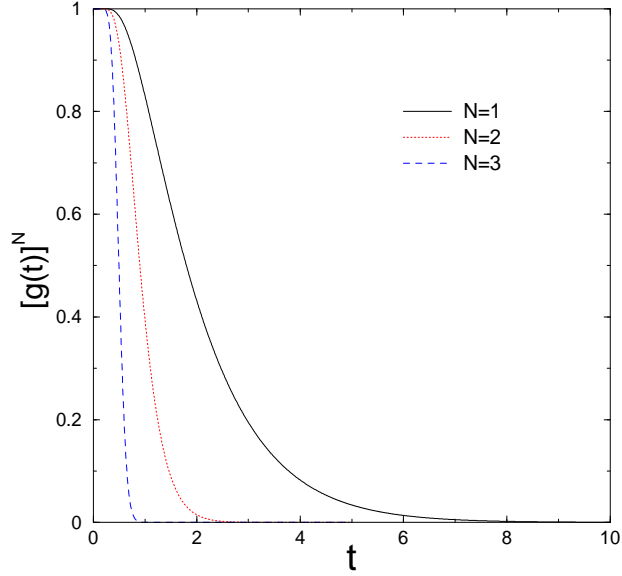}
\caption{\label{fig:gt} The function $[g(t)]^N$ plotted
as a function of $t$ for $N=1$ (black solid), $N=5$ (red dotted) and $N=50$
(blue dashed). 
As $N$ becomes
large, $[g(t)]^N$ almost approaches to a step function.}
\end{figure}

Next consider
the function $[g(t)]^N$ that appears in the integral in Eq. (\ref{bn3}).
As $N$ increases, the function $[g(t)]^N$ almost approaches
a step function (see Fig. ((\ref{fig:gt}))
\begin{equation}
[g(t)]^N \to \theta\left(t^*(N)-t\right)
\label{gt4}
\end{equation}
where the characteristic scale $t^*(N)$ decreases very slowly with increasing 
$N$. One can easily estimate $t^*(N)$ for large $N$ from the asymptotic
behavior in Eq. (\ref{gt3}). For small $t$, one finds, to leading order
for large $N$ 
\begin{equation}
[g(t)]^N \sim \left[1-2\, e^{-\pi^2/{4t}}\right]^N \simeq 
\exp\left[-2N\, e^{-\pi^2/{4t}}\right].
\label{gt5}
\end{equation}
Thus as $N$ increases, it approaches to $0$ rapidly for all
$t>t^*$ where $ 2N e^{-\pi^2/{4t^*}}\approx 1$. This provides
an estmaite of $t^*(N)$ which, to leading order for large $N$, reads
\begin{equation}
t^*(N) \simeq \frac{\pi^2}{4 \ln(N)}.
\label{gt6}
\end{equation}    

Therefore, for large $N$, using Eq. (\ref{gt4}), we can cut off the upper 
limit of the integral
in Eq. (\ref{bn3}) at $t=t^*(N)$ and replace 
$[g(t)]^{N-1}$ by $1$ over the interval
$t\in [0,t^*(N)]$.
Furthermore, over this small interval $t\in [0,t^*(N)]$, one can
replace the function $\sum_{n=1}^{\infty} n^2\, e^{-n^2\,t}$
by its small $t$ behavior $\simeq \sqrt{\pi}/{4t^{3/2}}$. 
Substituting these results in Eq. (\ref{bn3}) then yields, 
for large $N$, 
\begin{equation}
B_N(s) \simeq -\frac{N^2\, {\pi}^{N/2}}{2\Gamma(s)}\, \int_0^{t^*(N)} dt\, 
t^{s-(N+4)/2}\simeq -\frac{N^2 {\pi}^{N/2}}{2\Gamma(s)}\, 
\frac{[t^*(N)]^{s-N/2-1}}{(s-N/2-1)}.
\label{bn4}
\end{equation}
In deriving this result, we have assumed $s>(1+N/2)$.
After obtaining this large $N$ formula for $B_N(s)$,
we can now analytically continue it to $s\to 1$ which
finally yields
\begin{equation}
B_N=B_N(s\to 1)= N\, {\pi}^{N/2}\, [t^*(N)]^{-N/2}. 
\label{bn5}
\end{equation}
Upon using the expression for $t^*(N)$ from Eq. (\ref{gt6}) gives
the final large $N$ expression for $B_N$ in Eq. (\ref{ampN3}).


\begin{thebibliography}{999}
\bibitem{notation} Please note that 
in ref. \cite{MRZ1} the notation $q(z)$ was used for the hitting
probability (for exiting through $L$), but here we use $q_N(z)$ to denote the 
complimentary event of exiting the box through $0$. We apologise
if this causes any confusion.
\end{thebibliography}
\end{document}